\newif\ifprp
 \prptrue   % Preprint
\ifprp
 \documentstyle[twoside,aps,prb,epsf,floats]{revtex} 
\else
 \documentstyle[aps,prb,preprint]{revtex}
\fi

\begin{document}

\ifprp
\twocolumn[\hsize\textwidth\columnwidth\hsize\csname 
 @twocolumnfalse\endcsname 
\fi

\draft

\title{Nanoscale Field-Effect Transistors: An Ultimate Size Analysis}

\author{F. G. Pikus and K. K. Likharev}

\address{Department of Physics, State University of New York at Stony\\
Brook, Stony Brook, NY 11784-3800}

\date{\today}

\maketitle

\begin{abstract}
We have used a simple, analytically solvable model to analyze the characteristics
of dual-gate metal-oxide-semiconductor field-effect transistors (MOSFETs) with
10-nm-scale channel length $L$. The model assumes ballistic dynamics of 2D
electrons in an undoped channel between highly doped source and drain.
When applied
to silicon n-MOSFETs, calculations show that the voltage gain (necessary for
logic applications) drops sharply at $L\sim 10{\ {\rm nm}}$, while the
conductance modulation remains sufficient for memory applications until 
$L\sim 4{\ {\rm nm}}$.
\end{abstract}


\ifprp
 \vskip 2pc ] % end \twocolumn[...] 
\fi

The recent industrial forecast 
\cite{Roadmap} predicts that the progress 
in scaling down the metal-oxide-semiconductor field-effect
transistors (MOSFETs)
will 
result in MOSFETs with channel length $L$ at least as short as 70
nm. Such devices would allow the implementation of dynamic
random-access memories (DRAMs) with density up to $\sim 5{\rm \ Gb}/{\rm cm}%
^2$.
On the other hand, single-electron transistors (SET), 
in particular the recently
suggested SET/FET hybrids\cite{LikhKor95}, may allow room-temperature
operation of dynamic memories with a density of $100{\ {\rm Gb}}/{\rm cm}^2$
and beyond, using a silicon-based technology with minimum feature size below 
$\sim 5{\ {\rm nm}}$. Thus, it is very important to understand whether 
purely MOSFET-based devices can operate on a comparable scale. Recent
experiments \cite{Chou97} with $\sim$ 10-nm-long MOSFETs have shown
that their conductance can be gate-modulated by at least three orders of
magnitude. We are not aware, however, of any published analysis of $10-{\rm 
nm-}$ scale MOSFETs, with the exception of a couple of 
Monte-Carlo-calculated $I-V$ curves for some particular devices with $L=15$ nm 
\cite{Fukuma88} and $L=30$ nm \cite{Frank93}. 
The main goal of this work was
to use a simple model of nanoscale MOSFETs, which captures the essential
physics of such devices, for semi-analytical calculation of their basic
characteristics.
 All the
examples presented below are for silicon-based n-MOSFETs, although our
approach is certainly more general.

For nanoscale devices 
channel doping becomes {\em unacceptable}. In fact, the volume of
the channel region of a 10 nm-scale transistor is of the order of $2\cdot 10^{-19}{%
\ {\rm cm}}^3$, so that even doping at a level as high as $10^{20}{\ {\rm cm}%
}^3$ (beyond the degeneracy threshold for silicon) would result in
less than 20 dopants in the whole channel, and hence in large statistical
device-to-device variations of transistor parameters. 
For the same reason, contact doping has to be high 
($> 3\cdot10^{20} {\rm \ cm}^{-3}$). On the other hand,
channel doping is {\em unnecessary}, because the channel length is
comparable to the screening length
in highly doped source and drain, so that the carriers may be
delivered from the contacts\cite{Luryi87}. 
This is why in our model the
channel is a layer of an intrinsic semiconductor connecting $n^+$ 
source and drain (Fig. 1a). Because of the absence of impurities, electron
scattering is so small at our scale of channel length ($L\sim 10$ nm)
that it may be ignored\cite{Fukuma88,Frank93}, and the electron transport in the
channel considered as completely ballistic. Electrons in
the source and drain are assumed to be in thermal equilibrium.

Thickness $2s$ of the channel  is assumed to be so small that in fact it
presents a quantum well with 2D electron gas. For Si
(100) surface
the 6-fold valley degeneracy of
the bulk Si is lifted and only 2 valleys participate in the lateral
transport\cite{Ando}, unless the carrier density is so large that the electrons
start to populate higher subbands (which is not the case for the
devices
considered below).
The channel is sandwiched between plates of a ``dual gate'' \cite
{DualGateT,DualGateE}, the optimal structure to suppress the ``short channel
effects'' (for a review see, e.g., Ref.\onlinecite{Taur97}). 
If $2s$ is small enough, $z$-dependence of the
electric potential $\phi$ can be assumed quadratic inside the channel
and linear inside the dielectric (the well-known
parabolic approximation \cite{Young89}). Substituting this dependence
into the Poisson equation one arrives at 
a 1D equation for the distribution of electric potential $\phi$ at $z=0$
along the length of the device: 
\begin{equation}
\label{Poisson}
\frac{d^2\phi}{dx^2}-\frac{\phi -V_{\rm g}}{\lambda^2}=-\frac{4\pi\rho (x)}{
\kappa_1}. 
\end{equation}
Here $\rho =-en$ is the electric charge density averaged over the channel
thickness $2s$; within source and drain $\rho $ includes also the dopant
charge $+eN_{\rm D}.$ $V_{\rm g}$ is the gate-source voltage, and $\lambda $ is the
effective screening length

\begin{equation}
\label{ScrLength}\lambda ^2=\frac{s^2}2+\frac{\kappa_1}{\kappa_2}%
s(d-s). 
\end{equation}

\noindent For the examples presented below we took the dielectric 
constants to be
$\kappa_1=12$ (Si) and $\kappa_2=4$ (SiO$_2$).
The parabolic approximation is strictly valid\cite{Young89} if $\lambda$ is
much larger than $s$ but 
much less than relevant $x$-scales of the problem, notably, the
channel length and the screening length $\lambda_2$ of the electron gas without
the ground plane. $\lambda_2$ is close to the effective Bohr radius
$a_{\rm B} = \alpha \kappa \hbar^2/m e^2$, where $\alpha \approx 1.5$
accounts for valley degeneracy.

The second equation relating $\rho$ and $\phi$ follows from the condition
of conservation of the ballistic current components $j_\epsilon ^{\pm }$ for
each energy $\epsilon $: 
\begin{equation}
\label{Flux}j_\epsilon ^{\pm }(x)=-e\cdot n_\epsilon ^{\pm }(x)\cdot
v_\epsilon ^{\pm }(x)=const=j_{\epsilon 0}^{\pm }, 
\end{equation}
where $v^{\pm }(x)=\pm \sqrt{2[\epsilon - \Phi(x)]/m}$ is the $x$ -
component of the electron velocity, $\Phi(x) = -e\phi(x)$ is the
electron potential energy, and $j_{\epsilon 0}^{\pm }$ are currents
into the ballistic channel from the surfaces of source (sign $+$) and drain
(sign $-$). The latter currents can be found from the usual thermal equilibrium
distribution, then the velocity $v_\epsilon(x)$ and density $n_\epsilon(x)$ are found
with self-consistent values of $\Phi $. The total electron
density $n$ and current $j$ (per unit channel width) can the obtained as
integrals of $(n_\epsilon^+(x)+n_\epsilon^-(x))$ and $(j_{\epsilon
0}^+-j_{\epsilon 0}^-)$, respectively, over energies $\epsilon >
\Phi(x)$ (or $\epsilon > \Phi_{\max}$ if the potential maximum lies
between the point $x$ and the contact, Fig. 1b). For the parameters considered below, tunneling under
the potential barrier\cite{Tucker94} can be ignored for $L \ge 5$ nm.

The resulting simple set of equations allows the channel length $L$ and
current $j$ to be expressed explicitly as analytical (though bulky)
integrals for an arbitrary relation between temperature $T$ and Fermi
energy $\epsilon_{\rm F}$ in the source and drain, provided that $\Phi_{\max }$
is considered known (together with the real parameters of the system,
including the gate and source-drain voltages). The resulting function
$L=L(\Phi_{\max})$ can be numerically interpolated to the desired values of
channel length.

Figures 2-4 show the results of such semi-analytical calculations for a
Si/SiO$_2$ $n$-MOSFET with a contact doping level of $N_{\rm D}$=$3\cdot
10^{20}$ cm$^{-3}$, channel thickness $2s=1.5$ nm, and gate
oxide thickness $d-s=2.5$ nm. For these parameters,
the parabolic approximation is indeed applicable: Eq.~1 yields $\lambda
\approx
2.5$ nm, so that the ratio $\lambda/s$ is about 3.3.
Though the factor $\lambda/a_{\rm B}
\approx 1.2$ can be hardly considered as
large, we still believe that the parabolic approximation should work
reasonably well provided that the channel length is much larger than
$a_{\rm B}$, because the screening at $a_{\rm B}$
is polynomial\cite{Ando} (for a point charge, $\phi \sim r^{-3}$ at $r
\gg a_{\rm B}$), while that due to the gates is exponential.
Hence in a relatively long device ($L \gg
a_{\rm B}$) the wave vectors of the order of $a_{\rm B}^{-1}$ can give
only a small correction to our results, crudely equivalent to a channel
length uncertainty of the order of $a_{\rm B} \sim 2$ nm.

Figure 2 shows the distribution of electric
potential (solid lines) and electron density (dashed lines) along a
10-nm-channel MOSFET. Clearly, our model is capable of describing the
penetration of the electric field into source and drain, as well as the
pinch-off effect in relatively long devices (such as that shown in Fig. 2).
The $I-V$ curves of such devices show clear saturation (Fig. 3) and hence
relatively high voltage gain $G_{\rm V}=dV/dV_{\rm g}\mid_{I=const}$. Their
transconductance is also as high as that of the best present-day
transistors\cite{Taur97}, for example $S\approx 1.5$ S/mm
at $V=V_{\rm g}=0.5$ V for $L=10$ nm.

However, as the length $L$ becomes comparable to the effective screening
length $\lambda$, electron concentration in the channel becomes controlled 
more by the drain voltage and less by the gate.
The saturation disappears, and
the voltage gain above the threshold drops sharply, especially in the
range of $V_{\rm g}$ where the current is substantial (Fig. 4).
Beyond $\approx 10$ nm logic applications of
nano-MOSFETs become problematic. However, for DRAM applications (see, e.g., 
Ref.~\onlinecite{Prince91}) the voltage gain, as well as the depletion
operation mode of the transistors ($V_{\rm t} < 0$), 
are of minor importance, since the compact
memory cells can be controlled by drivers using more
conservative technology. What is really important for DRAMs is to have
the channel current modulated by at least 8 to 9 orders of magnitude (this determines
the necessary ratio of retention/refresh time to read/write time).

The modulation is limited from the side of small currents by two major effects
not accounted for in our model: the thermal activation of holes
and tunneling through the gate oxide. The hole activation (and hence the
sharp loss of gate control) takes place when the maximum potential in the
channel approaches the middle of the band gap; the corresponding region in Fig. 4
is coarsely hatched. To estimate the tunneling effects, we have
calculated the current taking into account the image charge effects and
using potential barrier height of $3.2$ eV and effective electron mass of
SiO$_2$ $0.4m_0$ 
(see, e.g., Ref. \onlinecite
{Brar 96}), and assuming that the gates overlap source and drain by 2
nm (the final conclusions are fairly insensitive to this value). Fine hatching in Fig. 4 shows the region where the tunneling
current becomes larger than the channel current. The boundary of this region
rises very rapidly with a decrease in oxide thickness; for $d-s=2$ nm
it corresponds to $j\sim 10^{-6}$ {\rm A/cm}. On the other hand, if the
oxide is thicker than $2.5$ nm, the transconductance and voltage gain
fall, while the modulation range remains virtually the same because of the
hole activation effects. Hence this value of oxide thickness may be
considered as optimal.
Thus the channel current modulation can hardly be better than
that shown in Fig. 4. One can see that the maximum modulation depth falls
below $3\cdot 10^8$ at $L\approx 4$ nm; crudely, this value may be
considered as the minimum length of silicon-based MOSFETs for application in
traditional DRAM cells. Of course, this does not preclude the possibility
that smaller transistors could be used in some novel memories based on
physical principles.

In our analysis, several other potentially important effects have been
neglected, including 
impact ionization, finite
rate of the energy relaxation, and source and drain resistance and
self-heating. Simple estimates show, however, that all these effects can be
ignored for our parameters. 

To summarize, we have 
found that for silicon-based MOSFETs room-temperature operation with
reasonable parameters is still feasible for channel length $L$ down to $\sim
10$ nm for logic circuits and $\sim 4$ nm for DRAM cells. Since we have
considered a virtually optimal MOSFET structure, it is hard to imagine that
these limits could be surpassed without suggesting some radically new
physical ideas. 

Fruitful discussions with D. Ferry, M. Fukuma, S. Luryi, V. Mitin, 
M. Shur, and J. Tucker are gratefully
appreciated. The work was supported in part by ONR/DARPA within the
framework of the Ultra Electronics program.

\begin{figure}
\ifprp
  \epsfxsize=3.5in
  \epsfysize=3in
  \epsffile{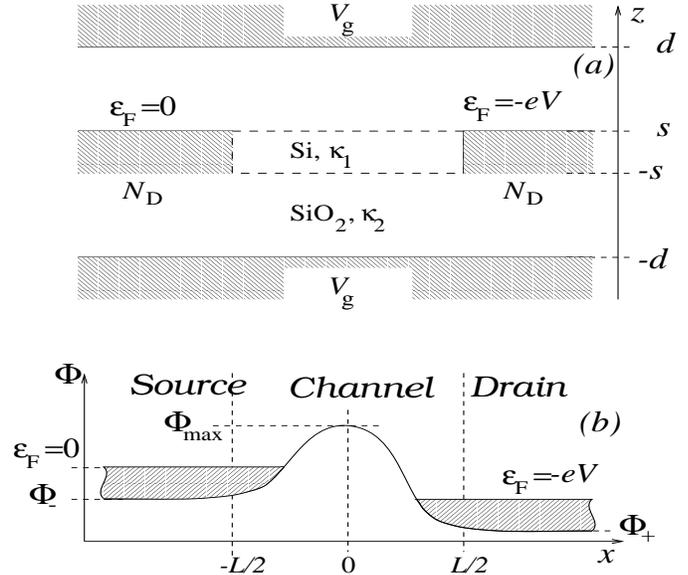}
\fi
\caption{
(a) Sketch of the MOSFET model considered in this work and (b) scheme of
the conduction band edge diagram for the electron potential energy
$\Phi=-e\phi$.
}
\end{figure}

\begin{figure}
\ifprp
  \epsfxsize=3.5in
  \epsfysize=3in
  \epsffile{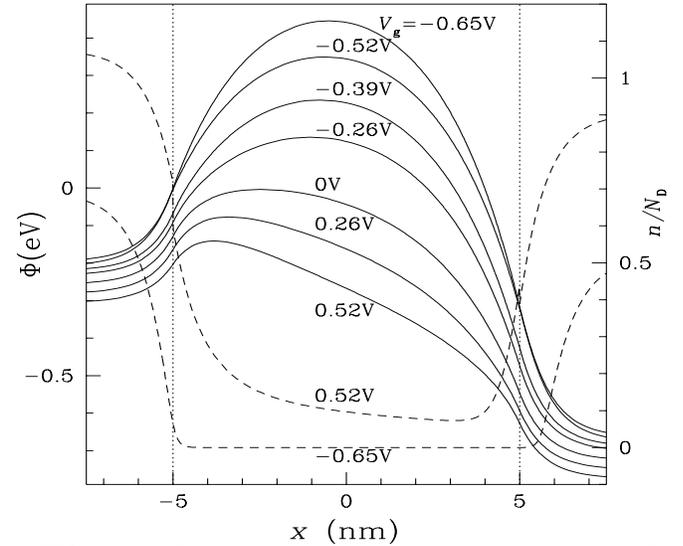}
\fi
\caption{
Distribution of electron potential energy $\Phi$ (solid lines) and density $n$ (dashed lines) along a 2D silicon n-MOSFET with a
10-nm-long intrinsic channel and $n^+$ source and drain 
($N_{\rm D}=3\cdot 10^{20}cm^{-3}$), for a moderate negative bias $V=-0.52V$ 
and several values of gate voltage $V_{\rm g}$. Geometric
parameters (Fig. 1a) are: $2s=1.5$ nm and $2d=6.5$ nm. Temperature $T=300$ K.
}
\end{figure}

\begin{figure}
\ifprp
  \epsfxsize=3.5in
  \epsfysize=3in
  \epsffile{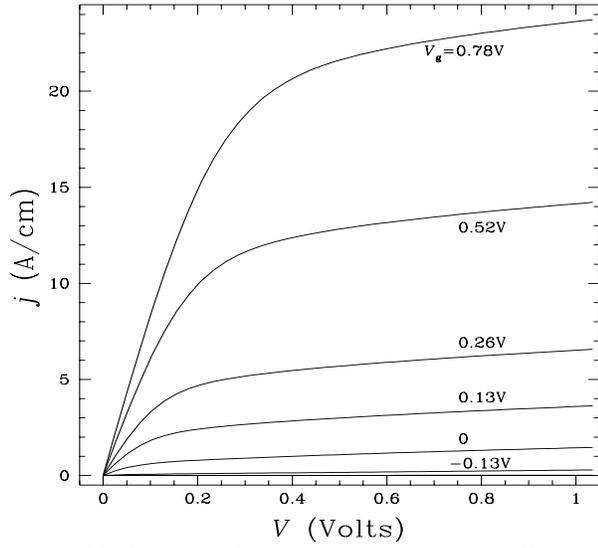}
\fi
\caption{Source-drain
$I-V$ curves for a n-MOSFET with $L=10$ nm
for several values of gate voltage $V_{\rm g}$. Device parameters are the
same as in Fig. 2.
}
\end{figure}

\begin{figure}
\ifprp
  \epsfxsize=3.5in
  \epsfysize=3in
  \epsffile{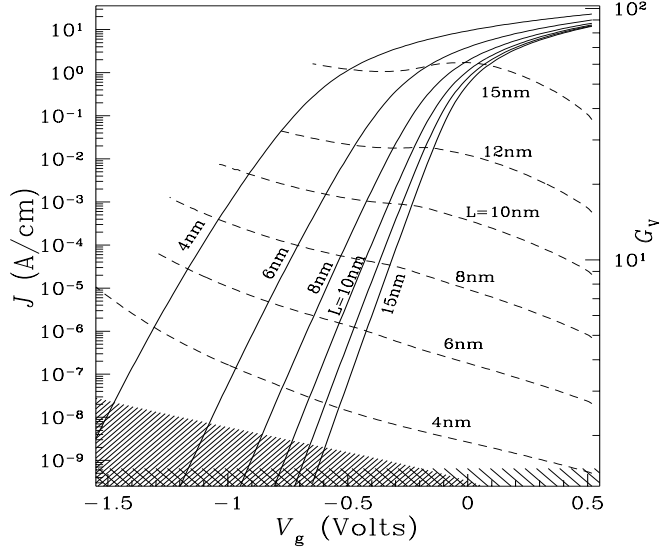}
\fi
\caption{
Linear current density $j$ (solid lines) and voltage gain 
$G_{\rm V}=dV/dV_{\rm g}\mid_{I=const}$ (dashed lines) as functions of gate voltage $V_{\rm g}$
for various channel lengths $L$ and source-drain voltage near the onset
of saturation ($V=0.52$ V). The fine hatching shows the area of parameters
where the gate leakage current exceeds the drain current. The coarse
hatching shows the region where the intrinsic carriers in the channel
cannot be ignored.
}
\end{figure}

\end{document}